\documentclass{article}

\usepackage{arxiv}

\usepackage[utf8]{inputenc} %
\usepackage[T1]{fontenc}    %
\usepackage{hyperref}       %
\usepackage{url}            %
\usepackage{booktabs}       %
\usepackage{amsfonts}       %
\usepackage{nicefrac}       %
\usepackage{microtype}      %
\usepackage{courier}
\usepackage{graphicx}
\usepackage[square,numbers]{natbib}
\usepackage{doi}

\usepackage{amssymb}
\usepackage{amsmath}
\usepackage{booktabs}
\usepackage{mathtools}
\usepackage{multirow}
\usepackage{cleveref}
\usepackage{svg}

\title{Navigating the Synthetic Realm: Harnessing Diffusion-based Models for Laparoscopic Text-to-Image Generation}

\date{November 30, 2023}	

\author{
    Simeon Allmendinger \\
    University of Bayreuth \\
    \texttt{simeon.allmendinger@uni-bayreuth.de} \\
    \AND
    Patrick Hemmer \\
    Karlsruhe Institute of Technology \\
    \texttt{patrick.hemmer@kit.edu}
    \And
    Moritz Queisner \\
    Charité – Universitätsmedizin Berlin \\
    \texttt{moritz.queisner@charite.de} 
    \And
    Igor Sauer \\
    Charité – Universitätsmedizin Berlin \\
    \texttt{igor.sauer@charite.de} 
    \And
    Leopold Müller \\
    University of Bayreuth \\
    \texttt{leopold.mueller@uni-bayreuth.de} 
    \And
    Johannes Jakubik \\
    Karlsruhe Institute of Technology \\
    \texttt{johannes.jakubik@kit.edu} 
    \And
    Michael Vössing \\
    Karlsruhe Institute of Technology \\
    \texttt{michael.voessing@kit.edu} 
    \And
    Niklas Kühl \\
    University of Bayreuth \\
    \texttt{kuehl@uni-bayreuth.de} 
}

\renewcommand{\shorttitle}{Navigating the Synthetic Realm}

\begin{document}

\maketitle
\begin{abstract}
Recent advances in synthetic imaging open up opportunities for obtaining additional data in the field of surgical imaging. This data can provide reliable supplements supporting surgical applications and decision-making through computer vision. Particularly the field of image-guided surgery, such as laparoscopic and robotic-assisted surgery, benefits strongly from synthetic image datasets and virtual surgical training methods. Our study presents an intuitive approach for generating synthetic laparoscopic images from short text prompts using diffusion-based generative models. We demonstrate the usage of state-of-the-art text-to-image architectures in the context of laparoscopic imaging with regard to the surgical removal of the gallbladder as an example. Results on fidelity and diversity demonstrate that diffusion-based models can acquire knowledge about the style and semantics in the field of image-guided surgery. A validation study with a human assessment survey underlines the realistic nature of our synthetic data, as medical personnel detects actual images in a pool with generated images causing a false-positive rate of 66\%. In addition, the investigation of a state-of-the-art machine learning model to recognize surgical actions indicates enhanced results when trained with additional generated images of up to 5.20\%. Overall, the achieved image quality contributes to the usage of computer-generated images in surgical applications and enhances its path to maturity.
\end{abstract}


\section{Introduction}
The development of Computer-Vision (CV)-enabled surgical applications requires large representative datasets. This is due to the complexity of adequately representing and addressing patient-specific variations in surgical data, such as anatomy and demographics, surgical skills and performances, as well as operating room setup such as surgical instruments and systems for acquiring data \cite{mascagni_computer_2022}. However, obtaining surgical data can be challenging due to privacy concerns, hospital regulations, cultural norms, and lacking technical capabilities \cite{mazer_video_2022}. Thus, state-of-the-art (SOTA) CV models can hardly be trained in the surgical domain. This demand for larger datasets motivates the usage of computer-generated synthetic images, as they offer new possibilities of augmenting images to create new data points \cite{chambon_adapting_2023}. 

Overall, there are two main application areas in which computer-generated images play a critical role. 
The first area is the use of Machine Learning (ML) models in the field of CV to perform detection or recognition tasks during surgery in real-time, e.g., tool detection \cite{kondo_lapformer_2021}. On the one hand, this can help surgeons perform complex surgeries by providing them with enhanced visualization and accurate identification of surgical targets. On the other hand, it allows standard surgical procedures to be performed faster and more safely. Thus, computer-generated data has the potential to enhance robustness and precision of surgical applications.

The second area in which computer-generated images are vital is the simulation of surgical procedures. Creating intervention-specific simulations of surgical situations would allow surgeons to safely plan and practice complex surgical procedures before performing them on patients. This can help reduce the risk of errors during surgery \cite{portelli_virtual_2020}. In addition, virtual surgical procedures can be used to train new surgeons \cite{jin_application_2021} enabling them to gain experience before performing real surgeries. Computer-generated images aid to create more realistic virtual surgical procedures faster compared to virtual models manually created by humans \cite{pfeiffer_generating_2019}.

In our work, we propose an approach to generate synthetic images (samples) in laparoscopic surgery from short text prompts by training a model to acquire knowledge about the style and semantics of a laparoscopic image. Style defines the appearance of the image, whereas semantics target the contextuality of its content. We choose text as input, as recent work shows that natural language is an excellent interface to enable non-technical users to interact with computer models \cite{wu2023visual}. On this basis, our work contributes threefold: First, the presence of style and semantics in our samples is underlined by the outcome of our survey. Participants with medical backgrounds caused a false-positive rate (FPR) of 66\%, when asked to detect actual images in an image pool with samples. Second, fidelity and diversity measures of our experiment indicate that the Imagen and Elucidated Imagen models are the preferred choice to generate samples. Third, the addition of samples increases the recognition average performance of a SOTA recognition model with up to 5.20\%. The code is made available\footnote{https://github.com/SimeonAllmendinger/SyntheticImageGeneration}.

\section{Related Work}

Former research in the domain of laparoscopic samples mostly utilizes image-to-image-translation on the basis of computer simulations \cite{pfeiffer_generating_2019}, phantoms \cite{sharan_mutually_2022}, segmentation maps \cite{marzullo_towards_2021} or simulated surgical tools \cite{colleoni_robotic_2021}; thus, the input of the models are images instead of text prompts. These authors present work addressing the lack of sufficient datasets relying on ground truth labels with Generative Adversarial Networks (GANs) by evaluating the value of samples in an ML downstream task. Marzullo et al. \citeyear{marzullo_towards_2021} also evaluate human assessment of samples in a survey.

Recent approaches also used text prompts as conditioning input, yet, they focus on gray-scale images in different medical imaging domains. Their studies adapt latent diffusion models to computer-generate lung X-rays \cite{chambon_adapting_2023} or magnetic resonance images \cite{pinaya_brain_2022} and human assessment or ML downstream tasks for evaluation purposes. Further work used the pre-trained Dall-e2 to create lung X-rays or produce colored dermatological images adding seed images as starting point to optimize a classifier \cite{ali_spot_2023,sagers_improving_2023}.

Research on the design of text-to-image models lately focuses on diffusion-based generative models \cite{karras_elucidating_2022,ramesh_hierarchical_2022,saharia_photorealistic_2022} exceeding the quality of GANs in certain situations \cite{dhariwal_diffusion_2021}. The performance of diffusion-based generative models is commonly determined by the Fréchet Inception Distance (FID) \cite{heusel_gans_2017}. Yet, recent work \cite{betzalel_study_2022} proposes improvements to the FID features investigating the Fréchet CLIP Distance (FCD) or enhances the FID presenting \textit{clean-fid} \cite{parmar_aliased_2022}.

Building on the advances in synthetic imaging, our work aims to bridge the gap between the medical and technical fields of research. Motivated by medical research with diffusion-based models, our work is the first to produce laparoscopic samples from text and derives conclusions on their value as a supplement for training a recognition model.

\section{Approach}
\label{sec:approach}

The goal of this study is to develop a diffusion-based text-to-image approach that can effectively learn both the style and semantics of laparoscopic images. Our approach is centered around the use of public laparoscopic datasets, which provide the necessary text-image pairs for training and testing. 

\begin{figure}[ht]
    \centering
    \includegraphics[width=\columnwidth]{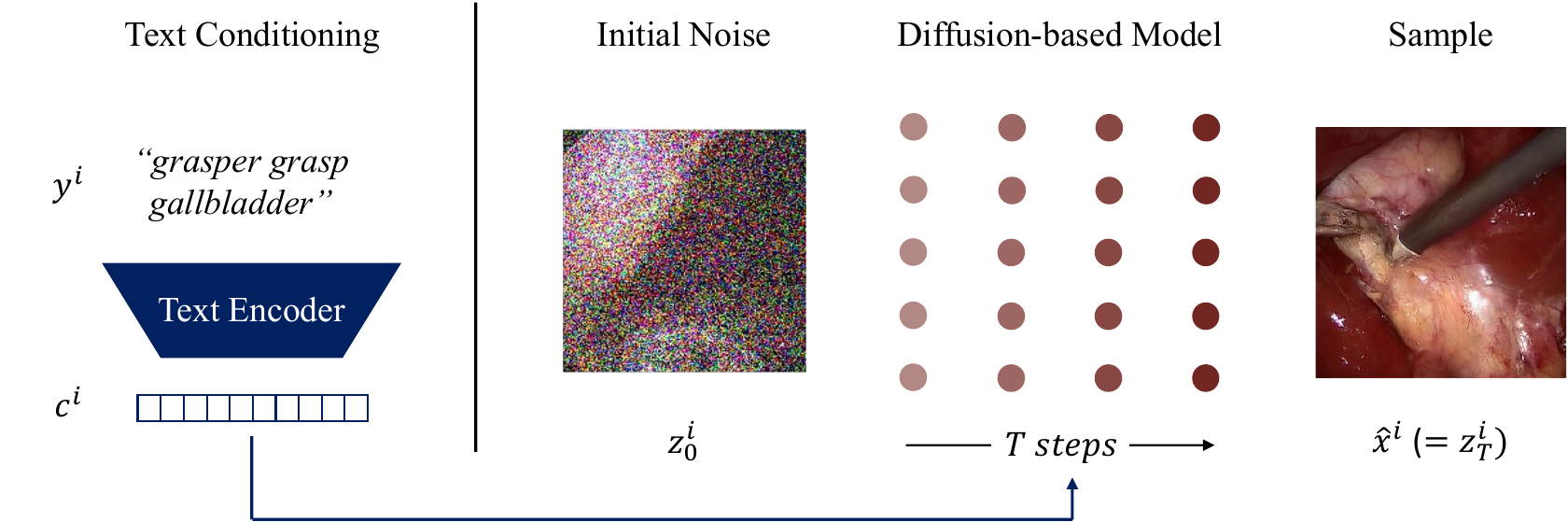}
    \caption{Schematic illustration of the text-to-image approach: Basic functioning of a diffusion-based model.}
    \label{fig:approach}
\end{figure}

At first, our approach shown in \Cref{fig:approach} minimizes the difference between a generated  sample \(\hat{x}^i\) conditioned on a text prompt \(y^i\) and the actual image \(x^i\) defining an objective function 

\begin{equation}
\label{eq:optimizationproblem}
    Q: min \space \sum_{i=1}^N \| \hat{x}^i(y^i) - x^i\|.
\end{equation}

We refer to the number of text-image pairs ($y \leftrightarrow x$) as \(N\) with data iterator $i$. In the second step, our text-to-image approach begins by pre-processing a text prompt \(y^i \), which is jointly drawn with the corresponding image \(x^i\) from the training dataset. In order to condition the sample generation of \(\hat{x} = \{\hat{x}^1, \hat{x}^2, ..., \hat{x}^N\}\) on \(y = \{y^1, y^2, ..., y^N\}\), the pre-processing calculates the corresponding text embeddings \(c = \{c^1, c^2, ..., c^N\}\) using the text encoders CLIP \cite{radford_learning_2021} or T5 \cite{raffel_exploring_2020} in accordance with the model design. Each text prompt \(y^i\) consists of concatenated strings of up to four categories in the following way:
\begin{equation}
    y^i = \underbrace{instrument + action + target}_\textit{triplet} + phase,
    \label{eq:text-prompt}
\end{equation}

\begin{figure}[ht]
    \centering
    \begin{tabular}{p{5cm} p{5cm} p{5cm} }
        \includegraphics[width=5cm]{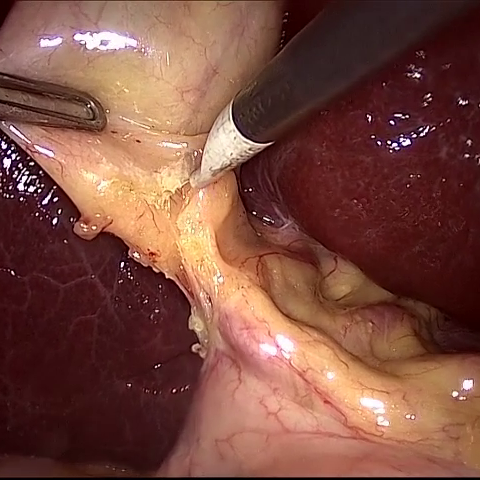} &
         \includegraphics[width=5cm]{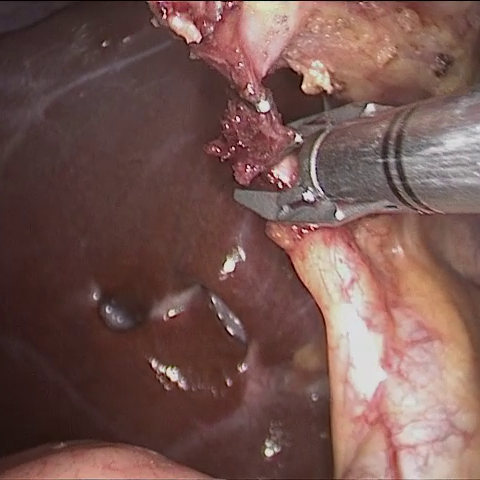} &
         \includegraphics[width=5cm]{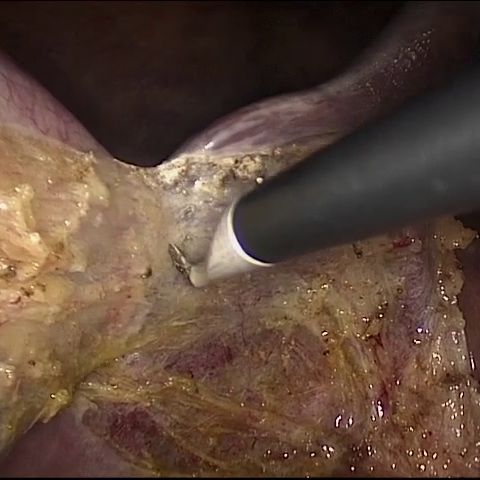} \\
        \small  (A) ``\textit{Hook and grasper grasp gallbladder in Calot Triangle Dissection}'' &
        \small (B) ``\textit{Clipper clip cystic pedicle in Clipping Cutting}'' &
        \small (C) ``\textit{Hook dissect omentum in Gallbladder Dissection}'' \\
        &&\\
        \includegraphics[width=5cm]{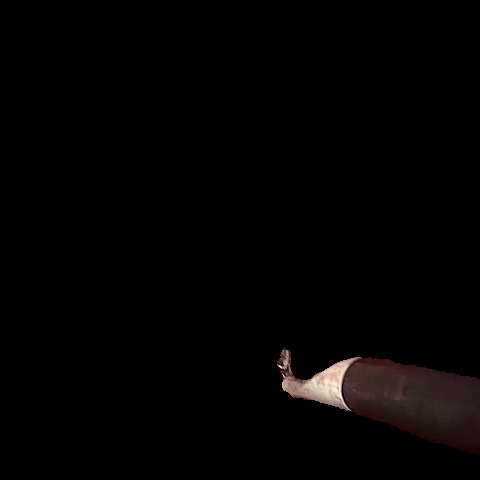}
        &
        \includegraphics[width=5cm]{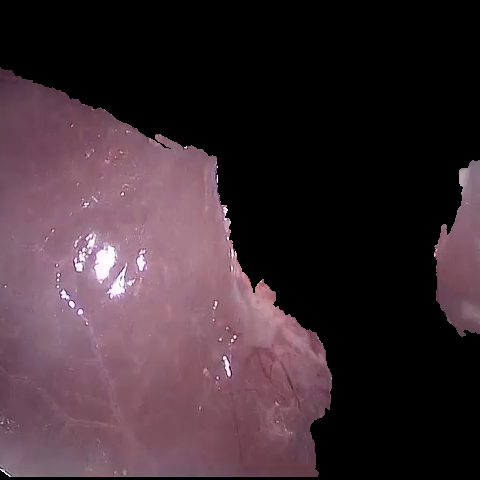} 
        &
        \includegraphics[width=5cm]{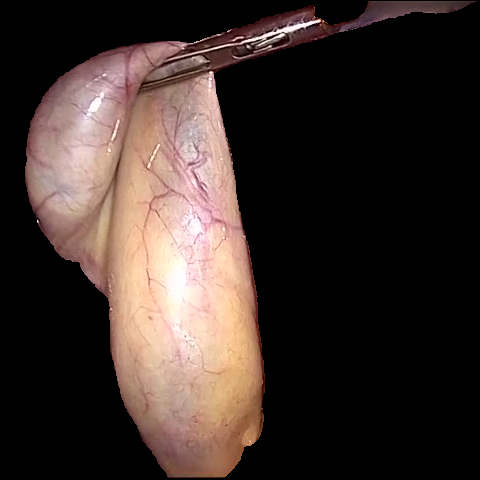} \\
        \small (I) ``\textit{hook''} &
        \small (II) ``\textit{liver''} &
        \small (III) ``\textit{gallbladder and grasper''}
    \end{tabular}
    \caption{Text-image pairs (A, B, C) of dataset CholecT45. Each image \(x^i\) is labeled with a text-prompt \(y^i\) (Eq. \ref{eq:text-prompt}). I, II and III show segmented data pairs of dataset CholecSeg8k.}
    \label{fig:text-image-pairs}
\end{figure}

An example for a text prompt is ``grasper grasp gallbladder in gallbladder dissection''. The images \(x\) are transformed for training using resizing, center cropping, and horizontal flipping. In addition, we create semantically segmented images aiming to support the specificity of the diffusion-based model. The images have the particularity that we cut out the regions of interest as individual images. While the background is set to be black, the regions visible are determined by semantic classes of surgical tools, organs, or anatomical structures. The corresponding text prompts comprise the class names of all classes displayed in the image (\Cref{fig:text-image-pairs}). 

In a third step, the pre-processed text-image pairs $y$ are fed into a diffusion-based generative model $f_{\theta}(y) = \hat{x}$ with trainable parameters \(\theta\). Diffusion models are latent variable models and have been widely studied in recent years due to their ability to generate high-quality samples from complex probability distributions \cite{karras_elucidating_2022}. They follow the idea to sequentially remove noise for \(T\) timesteps from an initial randomly distributed noise image \(z_0^i\) aiming to approximate the distributions \(z_T^i = \hat{x}^i \sim x^i\). Our work focuses on three diffusion-based generative models, which condition the noise removal on text: Dall-e2 \cite{ramesh_hierarchical_2022}, Imagen \cite{saharia_photorealistic_2022} and Elucidated Imagen \cite{karras_elucidating_2022}. As their code is not publicly accessible, we base our code on well-established implementations\footnote{https://github.com/lucidrains/DALLE2-pytorch}\footnote{https://github.com/lucidrains/imagen-pytorch}.

The Dall-e2 model is a generative neural network that decomposes the image generation process into a hierarchy of two stages. First, the authors used the Contrastive Language-Image Pre-Training (CLIP) model \cite{radford_learning_2021} to train a Prior with the objective to map a text embedding \(c_{text}^i\) from an initially given text prompt \( y^i \) to the corresponding image embedding latent \( c_{image}^i \). Second, \( c_{image}^i \) and \( z^i_0 \) are used in the Decoder to produce sample \(\hat{x}^i\). 

Imagen \cite{saharia_photorealistic_2022} consists of a cascaded diffusion model creating a pipeline, which includes U-Nets of different sizes. Since the authors proposed no Prior, they condition the denoising process directly on text embeddings \(c_{text}\). Imagen contains a number of new components: Frozen Text-To-Text Transfer Transformer (T5) \cite{raffel_exploring_2020}, dynamic thresholding to achieve higher photorealism and Efficient U-Nets, which converge faster. 

The Elucidated Imagen model bases on Imagen and a recent study \cite{karras_elucidating_2022}. This study investigated the idea of stochastic sampling, which delivered better results at the beginning of the training.

In addition to the models, we employ a Perception Prioritized (P2) weighting \cite{choi_perception_2022} for the Imagen and Elucidated Imagen models. It aids to recover signals from noisy data. Thus, P2 weighting suppresses weights for large Signal-to-Noise-Ratios occurring at the end of the denoising process, where the model learns imperceptible details. 

\section{Experiment}
\label{sec:experiment}

The evaluation of our text-to-image approach for generating laparoscopic images from text prompts involves two distinct outlets, each corresponding to the different application areas including the training of a ML image recognition task and the human perception of our samples as key to realistic surgical simulations. 

Initially, we aim to measure the added value arising from generated images for the ML recognition task of the Rendezvous (RDV) model \cite{nwoye_rendezvous_2022}. This is the winning model of the CholecTriplet Challenge at the MICCAI conference in 2022. The goal was to utilize ML techniques to enable real-time automated identification of 100 surgical actions in the form of text triplets, e.g. ``hook,dissect,omentum''. We use the RDV model as a baseline to evaluate the performance of our approach in generating high-quality laparoscopic images. Specifically, we assess whether the samples generated by our approach alter the performance of the model when trained with computer-generated images compared to training without them.

\begin{figure}[ht]%
\centering
    \begin{tabular}{p{6cm} p{6cm}}
        \includegraphics[width=6cm]{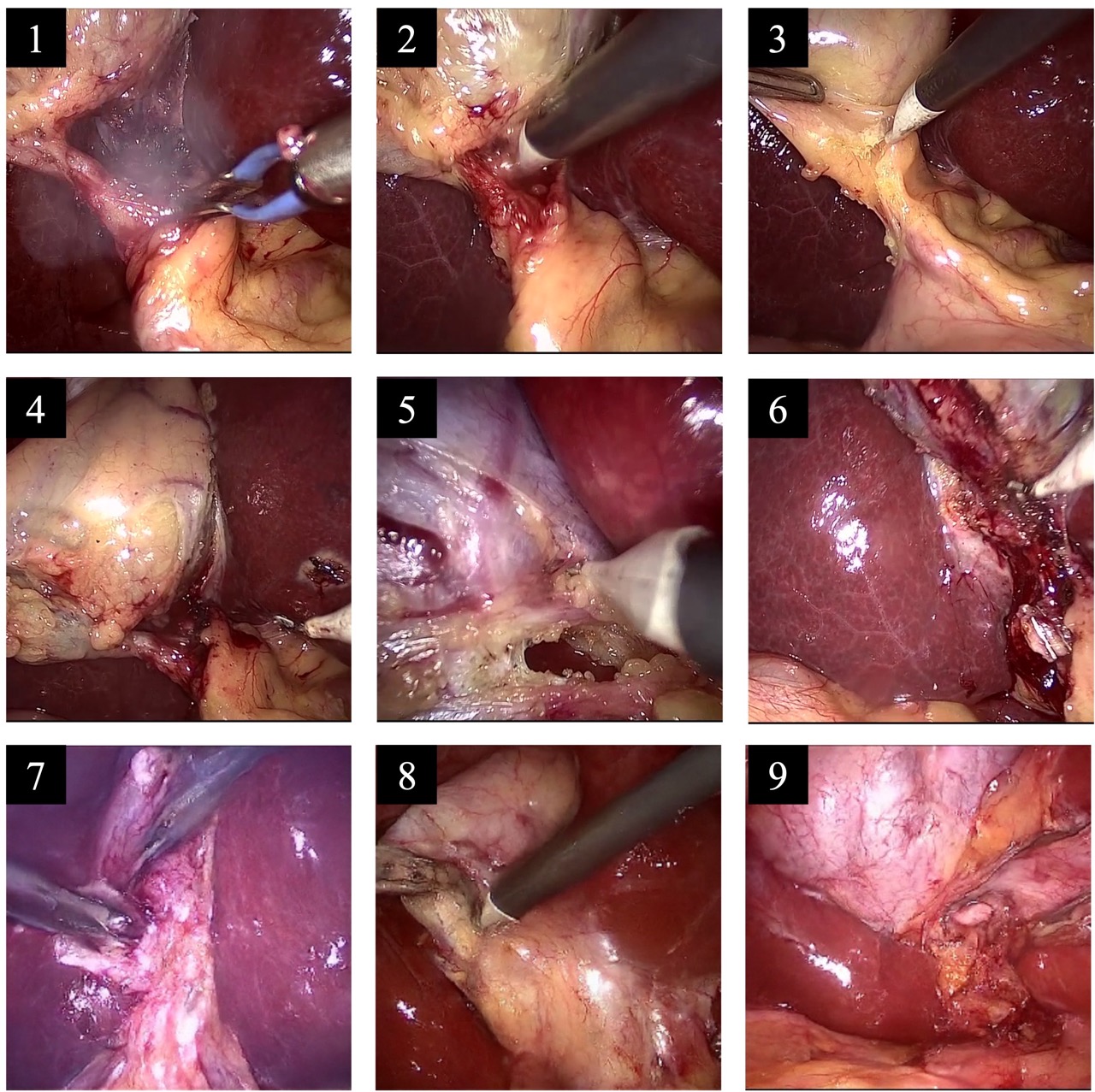} &
        \includegraphics[width=6cm]{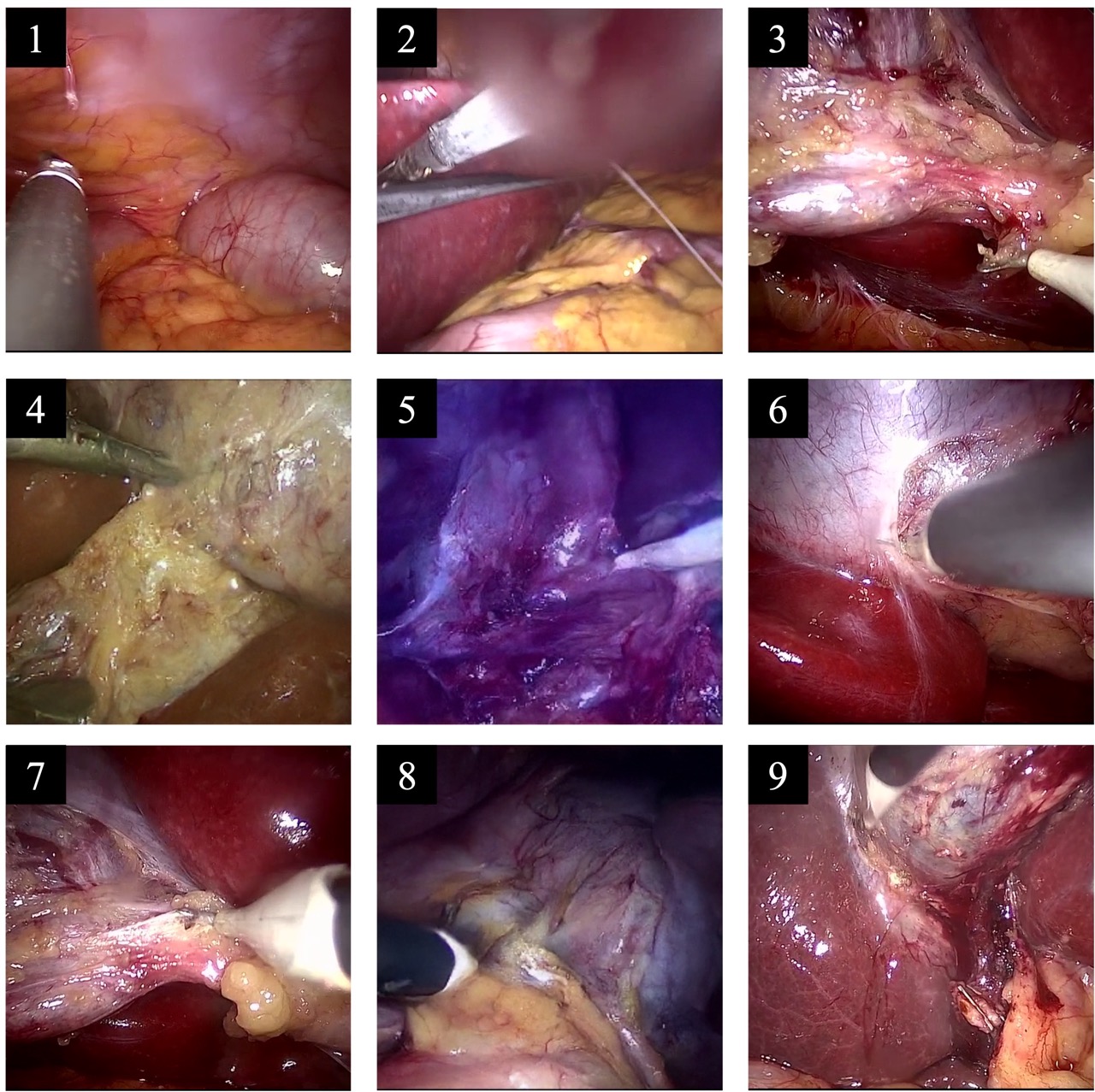} \\
    \end{tabular}
    \caption{\centering Two question design examples of the survey. Each question comprises an arbitrary mixture of nine real images and generated samples.}
    \label{fig:survey}
\end{figure}

In addition, we evaluate the authenticity of our samples through human assessment in a survey, which was approved by an institutional ethics committee. The participants were sourced from the prolific platform. The selection criteria included a medical profession, an age of at least 18 years, and proficient English. After agreeing to participate, individuals were asked to detect actual laparoscopic images in a set of 9 images, which comprises an arbitrary proportion of actual images and generated samples shown in \Cref{fig:survey}. The data pool contained 20 different questions with samples from the Imagen and Elucidated Imagen models respectively. Participants were asked 20 questions randomly and compensated with £9,07 per hour. We enrolled 24 participants between 26-67 years old, whereof 87.5\% identified as women, 12.5\% as men. On average, participants took 10:59 minutes and had to succeed an attention check. 

\subsection{Datasets}
Our work builds upon the datasets Cholec80 \cite{twinanda_endonet_2017}, CholecT45 \cite{nwoye_rendezvous_2022} and CholecSeg8k \cite{hong_cholecseg8k_2020}. The dataset Cholec80 contains recordings of 80 laparoscopic cholecystectomy procedures performed by 13 surgeons as videos. The videos are captured at 25 fps. Each frame is annotated with a label for both the seven surgical phases\footnote{\label{footnoteCholec80}\textit{Phases}: Preparation, Calot triangle dissection, Clipping and cutting, Gallbladder dissection, Gallbladder packaging, Cleaning and coagulation, Gallbladder retraction. \textit{Tools}: Grasper, Bipolar, Clipper, Scissors, Specimen bag, Irrigator, Hook.} and seven surgical tools\footnotemark[4]. The annotations were provided by a senior surgeon. The dataset CholecT45 is a subset of the Cholec80 with 45 videos. In addition, this dataset also includes annotations for 100 triplet classes in the form of \(triplet\):  \(< instrument, verb, target > \). The dataset CholecSeg8k is a publicly available dataset on Kaggle, that contains laparoscopic cholecystectomy images and their corresponding semantic segmentation maps with 13 labeling classes. It is also based on the Cholec80 and consists of 8,080 images. In our work, we obtain the images from the CholecT45 and the segmented images from the CholecSeg8k as $x$ (\Cref{fig:text-image-pairs}). The labels $y$ are derived using the triplet annotations from CholecT45 and phase labels from Cholec80.

\subsection{Setup}
The diffusion-based models are based on pre-trained text encoders. For the Dall-e2 model we use the English CLIP model \(ViT-B-32\) (laion2B-s34B-b79K)\footnote{https://github.com/mlfoundations/open\_clip}. For the Imagen and Elucidated Imagen model, we implement the T5 model \(t5-v1_1-base\)\footnote{https://huggingface.co/google/t5-v1\_1-base}. For all models, we apply various conditioning scales $\psi$ investigating the weight of text conditioning on our samples. The hardware consisted of up to 35 NVIDIA Tesla V100 GPUs. 

\subsection{Metrics}

Building upon previous work \cite{betzalel_study_2022}, we determine fidelity and diversity as the key measures. Fidelity refers to the realness of each input, whereas the degree of diversity encompasses how well samples replicate the variances present in real images \cite{naeem_reliable_2020}.

To quantitatively assess the fidelity of the generated images, we use four metrics: FID \cite{heusel_gans_2017}, \textit{clean-fid} \cite{parmar_aliased_2022}, FCD \cite{betzalel_study_2022} and Kernel Inception Distance (KID) \cite{binkowski_demystifying_2023} with 10 subsets containing 1,000 images. The diversity of the images is visually addressed with the T-distributed Stochastic Neighbor Embedding (TSNE) on the basis of ResNet50 embeddings. 

We evaluate the value of our samples for the RDV model with the Recognition Average Precision (RAP) following the proposed metrics and CholecT45 data splits \cite{nwoye_data_2023}. For the survey, we apply a confusion matrix, FPRs and true-positive rates (TPRs). 

\section{Results}\label{sec:results}

\begin{figure}[ht]%
    \centering
    \includegraphics[width=15cm]{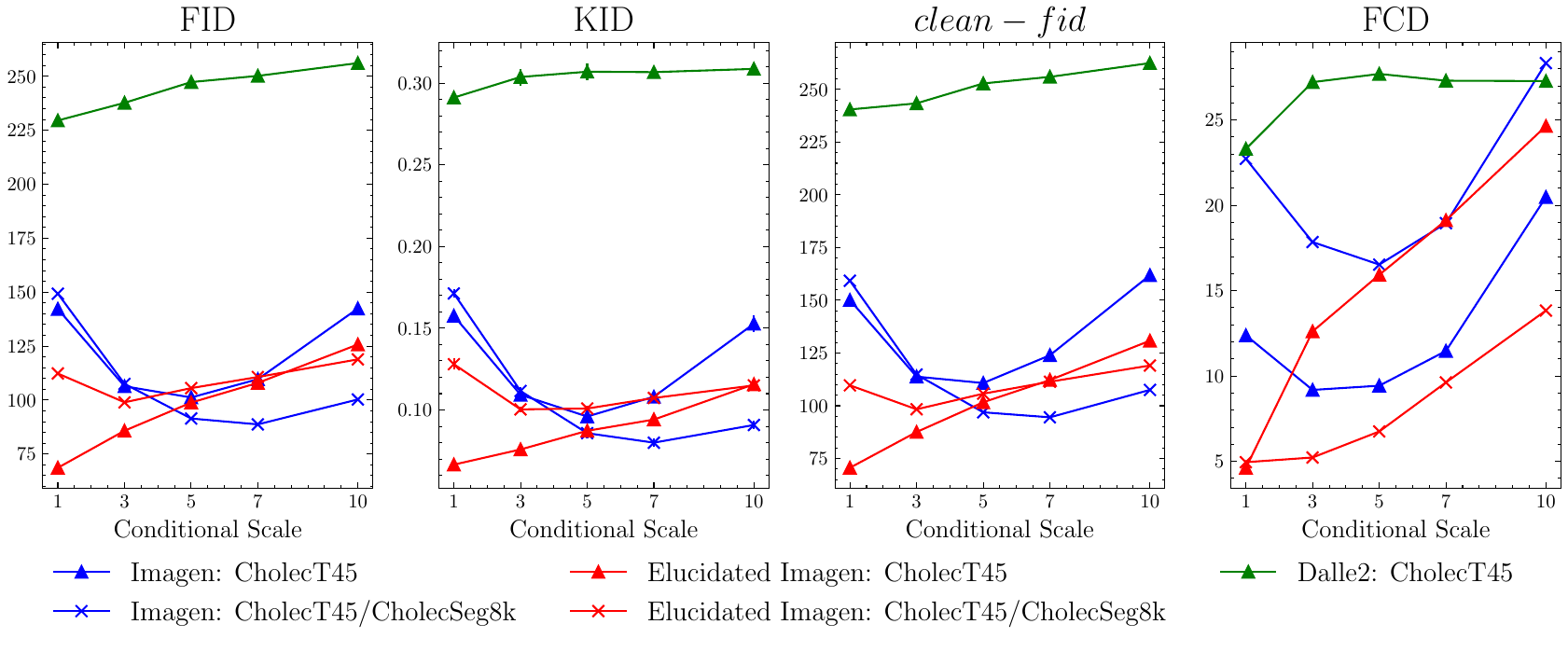}
    \caption{\centering Scores of FID, KID, \textit{clean-fid}, FCD; 10,000 real images and samples respectively for each measurement; models trained on CholecT45 (and CholecSeg8k).}
    \label{fig:imagenmetrics}
\end{figure}

We analyzed our text-to-image approach according to fidelity and diversity. At first, \Cref{fig:imagenmetrics} displays the scores of the metrics FID, KID, \textit{clean-fid}, and FCD. The graphs differentiate between conditioning scales $\psi$ of 1, 3, 5, 7, and 10 as well as the usage of the additional segmented images from the dataset CholecSeg8k. For each configuration, 10,000 samples were produced from random text-prompts taking up to 10h. The findings indicate that the quality of samples varies across different models, conditioning scales, and the additional segmented images of the dataset CholecSeg8k. In addition, the FID and FCD metrics exhibit significant differences in magnitude and divergent relative scores. Notably, the Elucidated Imagen model performs best in terms of these metrics at low conditioning scales. However, the addition of the segmented images of the dataset CholecSeg8k produces ambiguous outcomes, especially considering the FID and FCD for the Elucidated Imagen model.

\begin{figure}[ht]
    \centering
    \begin{tabular}{p{7.5cm} p{7.5cm}}
        \includegraphics[width=7.5cm]{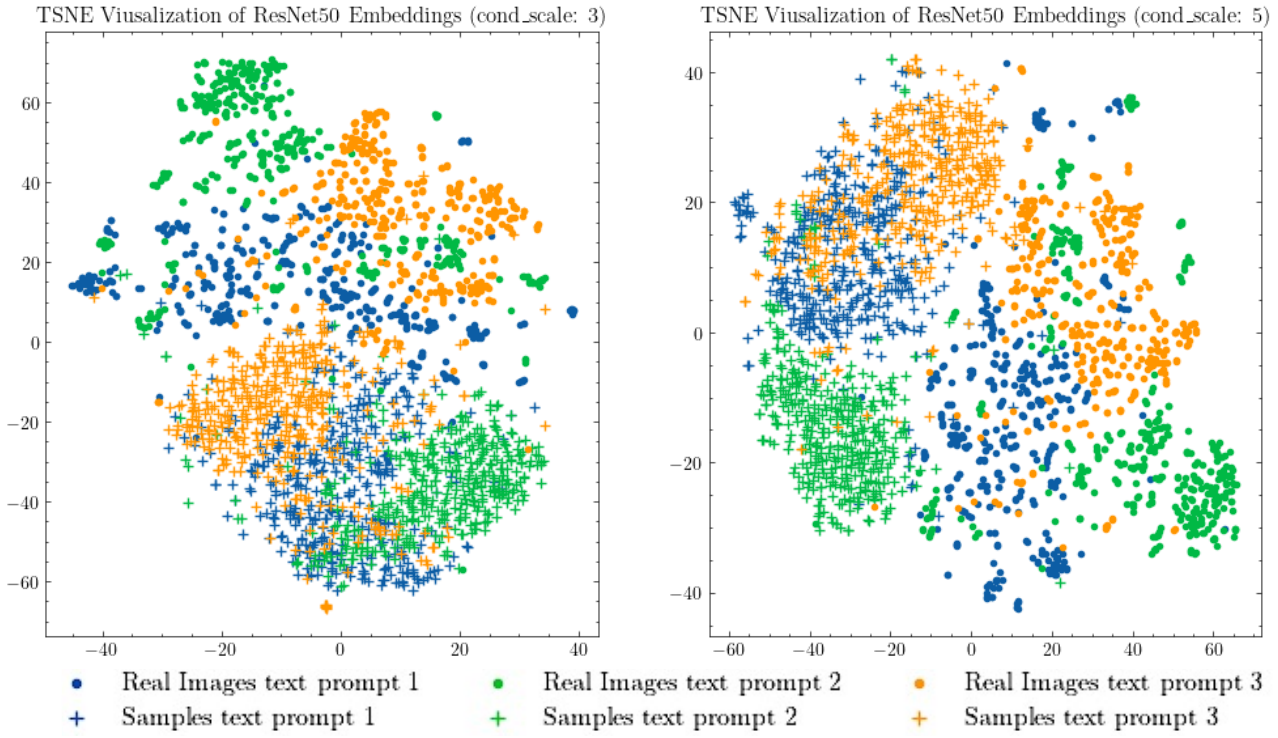}
        &
        \includegraphics[width=7.5cm]{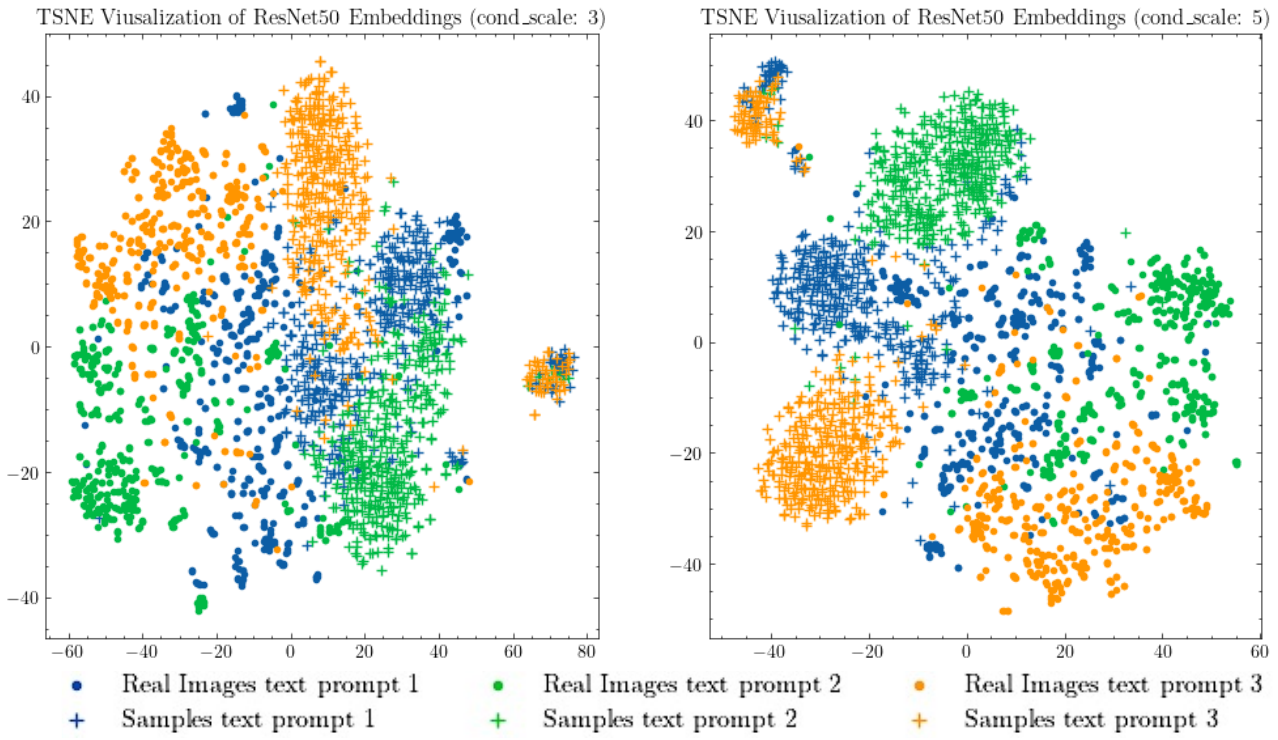} 
        \\
        \small \centering (a) Imagen model
        &
        \small \centering (b) Elucidated Imagen model
    \end{tabular}
    \caption{TSNE of real image and sample embeddings (Number of components: 2; perplexity: 30; number of images: 1200; 3 text prompts: \textit{``(1) grasper retract liver in gallbladder dissection'', ``(2) hook dissect gallbladder in gallbladder dissection''}, and \textit{``(3) grasper retract gallbladder in calot triangle dissection''. Generated images show a similar bias.}}
    \label{fig:tsne}
\end{figure}

The diversity of the images is evaluated by the TSNE distributions of 1,200 random actual image embeddings and sample embeddings respectively from three text prompts. The Imagen and the Elucidated Imagen model both trained on the dataset CholecT45 are separately shown in \Cref{fig:tsne}a and \Cref{fig:tsne}b considering a conditioning scale $\psi$ of 3 and 5. The figures allow for three observations. First, if compared to the actual images, both figures emphasize that the samples show different characteristics and are not solely a duplicate of existing images. Furthermore, all embeddings of the samples present a similar bias. Second, the sample embeddings appear in clusters, which roughly align with the text prompt conditioning. However, the clusters show overlaps. Interestingly, the actual images from different text-prompts seem to have more similar characteristics. Third, the model choice and the conditioning scale have an effect on the diversity of the samples. Both models reduce overlapping with a higher conditioning scale $\psi$ of 5, while the Elucidated Imagen model appears to have a better approach in producing similar samples from equal text prompts.

We qualitatively address the models in \Cref{table:synthetic_images_comparison} by comparing the produced samples conditioned on the text prompt: \textit{``grasper retract gallbladder in calot triangle disscetion''}. It can be observed that with increasing $\psi$ the sensitivity to noise rises. Another noteworthy observation is that low conditioning on text frequently results in the absence of certain components of the text prompt in the generated sample, e.g., lack of surgical tools. Moreover, as indicated by the fidelity, the Dall-e2 model with our configuration is not capable to produce realistic samples. The style of these samples is over-saturated and the context is not easily recognizable. 

\begin{table*}[h!]
\centering
\resizebox{0.9\linewidth}{!}{%
    \begin{tabular}{l || c | c | c ||| c}
        \toprule
        \multirow{2}{*}{$\psi$} & \multirow{2}{*}{Dall-e2} & \multirow{2}{*}{Imagen} & \multirow{2}{*}{Elucidated Imagen} & Elucidated Imagen +\\
        & & & & CholecSeg8k \\
        \midrule
        &
        \multirow{2}{*}{\includegraphics[width=0.2\textwidth]{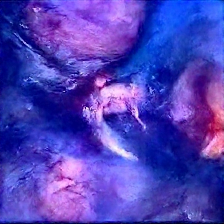}}&
        \multirow{2}{*}{\includegraphics[width=0.2\textwidth]{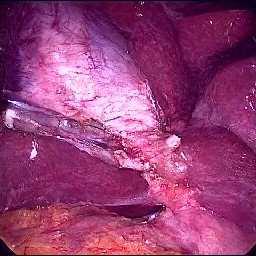}}&
        \multirow{2}{*}{\includegraphics[width=0.2\textwidth]{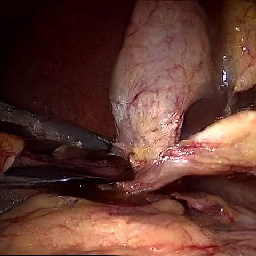}}&
        \multirow{2}{*}{\includegraphics[width=0.2\textwidth]{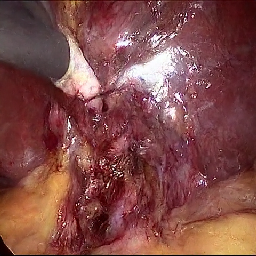}}\\
        &&&\\ &&&\\ &&&\\ 1&&&\\ &&&\\ &&&\\ &&&\\ &&&\\ &&&\\
        &
        \multirow{2}{*}{\includegraphics[width=0.2\textwidth]{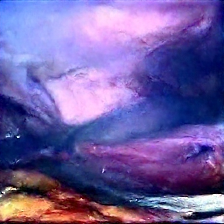}}&
        \multirow{2}{*}{\includegraphics[width=0.2\textwidth]{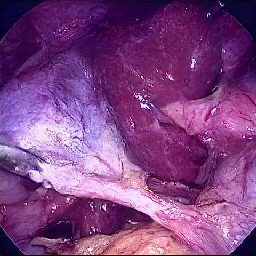}}&
        \multirow{2}{*}{\includegraphics[width=0.2\textwidth]{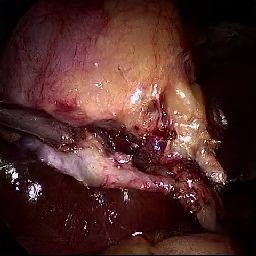}}&
        \multirow{2}{*}{\includegraphics[width=0.2\textwidth]{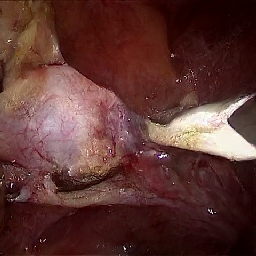}} \\
        &&&\\ &&&\\ &&&\\ 3&&&\\ &&&\\ &&&\\ &&&\\ &&&\\ &&&\\  

        &
        \multirow{2}{*}{\includegraphics[width=0.2\textwidth]{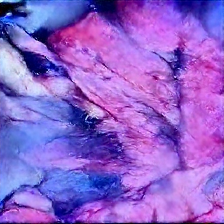}}&
        \multirow{2}{*}{\includegraphics[width=0.2\textwidth]{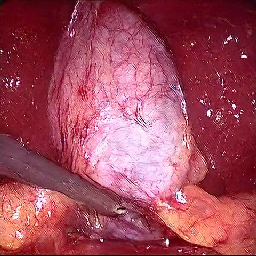}}&
        \multirow{2}{*}{\includegraphics[width=0.2\textwidth]{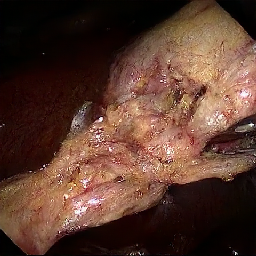}}&
        \multirow{2}{*}{\includegraphics[width=0.2\textwidth]{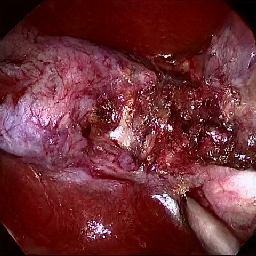}}\\
        &&&\\ &&&\\ &&&\\ 5&&&\\ &&&\\ &&&\\ &&&\\ &&&\\ &&&\\ 
        
        &
        \multirow{2}{*}{\includegraphics[width=0.2\textwidth]{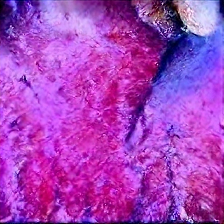}}&
        \multirow{2}{*}{\includegraphics[width=0.2\textwidth]{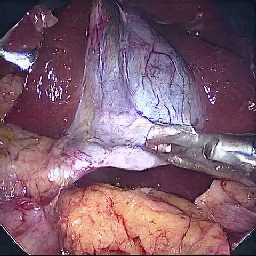}}&
        \multirow{2}{*}{\includegraphics[width=0.2\textwidth]{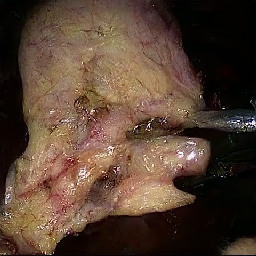}}&
        \multirow{2}{*}{\includegraphics[width=0.2\textwidth]{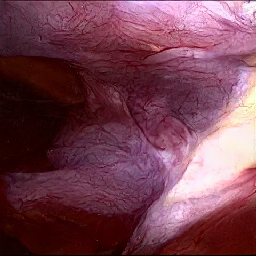}}\\
        &&&\\ &&&\\ &&&\\ 7&&&\\ &&&\\ &&&\\ &&&\\ &&&\\ &&&\\ 

        &
        \multirow{2}{*}{\includegraphics[width=0.2\textwidth]{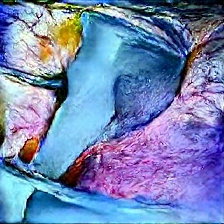}}&
        \multirow{2}{*}{\includegraphics[width=0.2\textwidth]{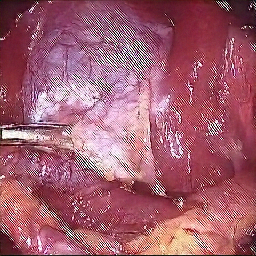}}&
        \multirow{2}{*}{\includegraphics[width=0.2\textwidth]{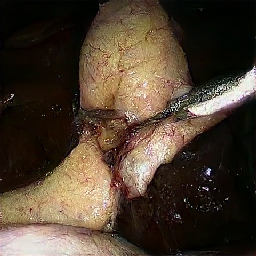}}&
        \multirow{2}{*}{\includegraphics[width=0.2\textwidth]{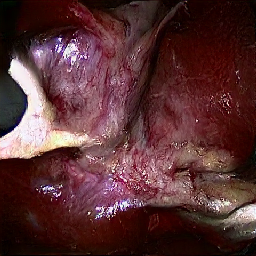}}\\
        &&&\\ &&&\\ &&&\\ 10&&&\\ &&&\\ &&&\\ &&&\\ &&&\\ &&&\\

        \bottomrule
    \end{tabular}}
\caption{On the left, the table displays samples conditioned on the text prompt \textit{``grasper retract gallbladder in calot triangle disscetion''} from Dall-e2, Imagen and Elucidated Imagen model applying five conditioning scales $\psi$. Moreover, five generated images from the Elucidated Imagen model are provided, additionally trained on the segmented images of the dataset CholecSeg8k using various conditional prompts: (1) \textit{``hook dissect gallbladder in calot triangle dissection''}; (3) \textit{``grasper retract gallbladder and hook dissect gallbladder in calot triangle dissection''}; (5) \textit{``grasper retract gallbladder in clipping cutting''}; (7) \textit{``bipolar coagulate abdominal wall cavity in gallbladder extraction''}; (10) \textit{``grasper retract gallbladder and hook dissect cystic duct in calot triangle dissection''}.}
\label{table:synthetic_images_comparison}
\end{table*}

\subsection{Evaluation of ML Recognition Task}
\label{subsec:machine_learning_evaluation}

\begin{figure}[hb]
    \centering
    \includegraphics[width=14cm]{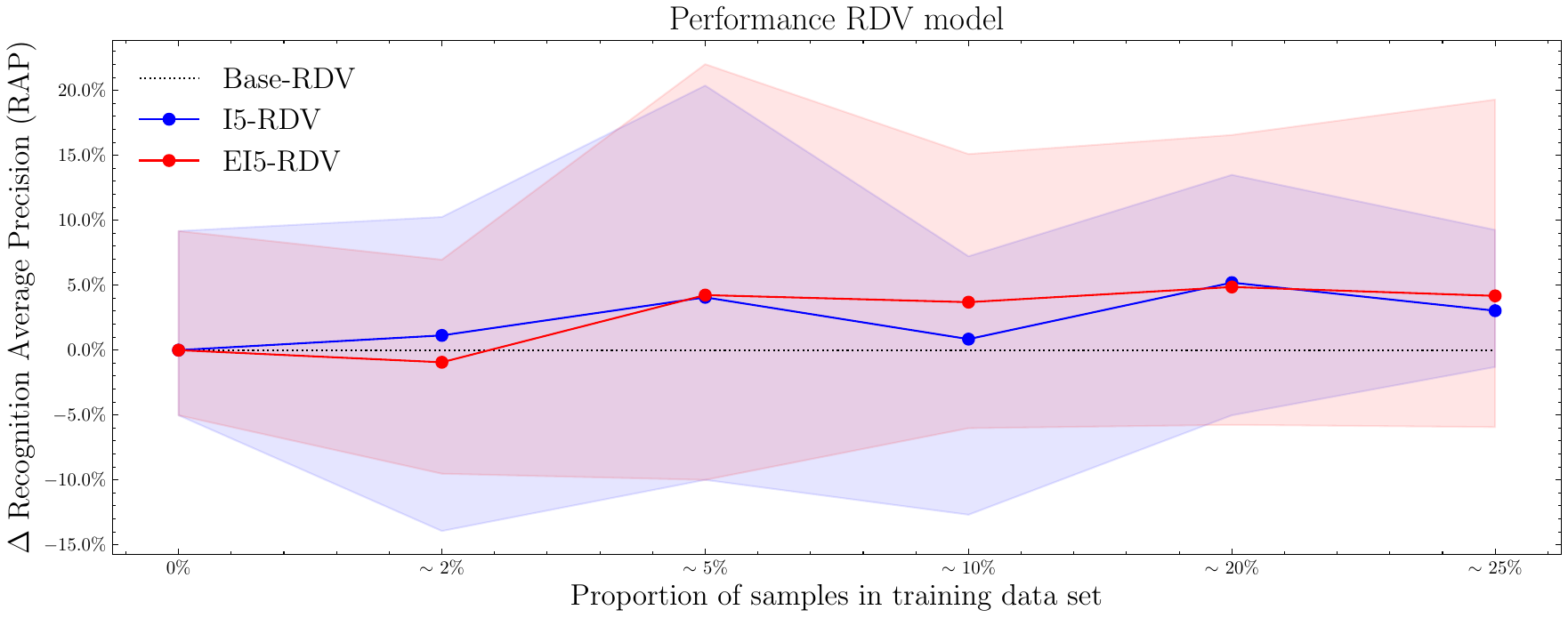}
    \caption{RDV model: $\Delta\%$ of RAP for various proportions of samples (data point $\widehat{=}$ 50 runs averaging over a 5-fold cross-validation; bands $\widehat{=}$ min/max score; epochs $=$ 100). Including synthetic samples in the training data improves the ML performance on the recognition task.}
    \label{fig:rendevouznetworkresults}
\end{figure}

\Cref{fig:rendevouznetworkresults} contains the results of the RAP for the RDV model. We distinguish between a Base-RDV model trained on the dataset CholecT45 achieving an initial performance of 23.11\% RAP as well as the I5-RDV and EI5-RDV models, which were additionally trained using samples created by the Imagen model and the Elucidated Imagen model. Both apply a conditioning scale $\psi$ of 5. The sample datasets consist of 3,000 up to 30,000 generated images from three text prompts: \textit{``(1) grasper retract liver in gallbladder dissection'', ``(2) hook dissect gallbladder in gallbladder dissection''}, and \textit{``(3) grasper retract gallbladder in calot triangle dissection''}. These samples were annotated with the corresponding text-based triplet labels: Triplet A: \textit{"grasper,retract,liver"}; Triplet B: \textit{"hook,dissect,gallbladder"}; Triplet C: \textit{"grasper,retract,gallbladder"}. Overall, the samples represent a proportion of 3.75-75\%, 1.75-35\%, and 1-20\% in their respective triplet classes. The RAP scores reveal three findings. First, our samples have, on average, a positive effect on the RAP with I5-RDV and EI5-RDV exceeding the Base-RDV despite high variability by up to 5.20\% and 4.86\%. This suggests that the samples contain valuable visual augmentations. Second, the scores also indicate that variances of the RAP increase. Third, a proportion of 5\% of synthetic samples in the training data already includes the majority of synthetic benefits for the I5-RDV and EI5-RDV.

\begin{table}[ht]
    \centering
    \begin{tabular}{l | p{1.5cm} | p{1.5cm} | p{1.5cm} | p{1.5cm} }
    \toprule
    Model & \multicolumn{2}{c |}{Imagen} & \multicolumn{2}{c}{Elucidated Imagen}\\
    \midrule
    & \multicolumn{1}{c |}{\textit{True}} & \multicolumn{1}{c |}{\textit{False}} &  \multicolumn{1}{c |}{\textit{True}} &  \multicolumn{1}{c}{\textit{False}}\\
    \textit{Positive}  & \multicolumn{1}{c |}{341}  & \multicolumn{1}{c |}{492} & \multicolumn{1}{c |}{314}  &  \multicolumn{1}{c}{425} \\
    \textit{Negative} &  \multicolumn{1}{c |}{253} &  \multicolumn{1}{c |}{354} & \multicolumn{1}{c |}{271}  & \multicolumn{1}{c}{430} \\
    \midrule
    TPR & \multicolumn{2}{c |}{49.07\%} & \multicolumn{2}{c}{\bf{42.20\%}} \\
    FPR & \multicolumn{2}{c |}{\bf{66.04\%}} & \multicolumn{2}{c}{61.06\%} \\
    \bottomrule
    \end{tabular}
    \caption{Survey results reveal that participants were challenged detecting the actual laparoscopic images.}
    \label{table:survey}
\end{table}

\subsection{Evaluation of Survey}
\label{subsec:survey_evaluation}
The results of the survey (\Cref{table:survey}) yield a FPR of 66.04\% (Imagen) and 61.06\% (Elucidated Imagen) targeting the ability of the participants to detect the actual laparoscopic images. Thus, the samples of the Imagen model appeared more real. However, the TPRs of 49.07\% (Imagen) and 42.20\% (Elucidated Imagen) highlight that participants performed less well detecting the actual images among samples of the Elucidated Imagen model. Overall, the medical participants were highly challenged by the realistic characteristics of the samples, which validates their authenticity. 

\subsection{Path to Deployment}
\label{subsec:pathtodeployment}
The training of SOTA CV models requires extensive annotated datasets, which are usually not present in surgery. Our findings emphasize that samples from diffusion-based models can not only be a valuable addition to existing datatsets, but are also already annotated. To apply our approach, we propose to split existing datasets into training, validation, and test data. The former builds the basis to train a diffusion-based model, which is tuned using the validation data. The test data finally evaluates the performance of the application.

Another way of deployment is the provision of our approach in order to enable surgeons to generate surgical context with their own descriptions. Furthermore, since text embedding is one form of conditioning, the development of surgical applications can extend this approach. Interactive simulations become possible, as agents can condition the next frame of a virtual scene on various inputs, such as text, current frame or position of the surgical tools.

\section{Limitations and Societal Impact}
\label{sec:limitationssocietalimpact}

Our findings underline the potential of diffusion-based samples to be a reliable supplement for CV-enabled surgical applications. Particularly, the positive effect of samples included in the training of the RDV model proves the added value of our approach. It can be argued that solely one recognition model is investigated, however, the RDV model is one of the most advanced and extensive recognition models with 100 different classes in the laparoscopic domain; thus, the complexity is rather high. In addition, the human experienced realism of the samples builds the basis for synthetic visual data for surgical training purposes. Since human assessment highly reacts to small distortions in samples, the positive results deserve special attention.

The bias in diversity between actual images and samples suggests that the generation of synthetic data still remains a persistent challenge. While more precise text prompts could improve sample quality in our approach, successful outcomes in this field hinge on a larger edited data pool, as existing models are limited in their ability to create new images from style or semantics not included in the initial dataset. Therefore, it is essential to avoid a positive feedback loop in which generative models are partially trained on samples, as this risks exacerbating the bias observed in our diversity analysis. Such a loop could trigger a self-reinforcing dynamic that impairs model performance on real-world data by fostering over-reliance on synthetic data. To mitigate bias and improve generalizability, generative models must be initially trained on representative real-world datasets.

Recent work \cite{carlini_extracting_2023} revealed that diffusion-based models tend to remember specific images from their training data and reproduce them during generation. Their findings suggest that these models are less private compared to previous generative models. This lack of privacy must be considered with regard to legal issues and concerns of patients. Therefore, addressing these vulnerabilities may require new approaches to privacy-preserving training. 

An advantage of our approach is the text-based interface, which makes the generation more transparent and easily accessible while improving a \textit{global} understanding of the generative behavior of the model. However, the investigated models must be further addressed regarding \textit{local} explainability in particular, which aims to understand the direct influence of single features (e.g., words) on the sample \cite{markus_role_2021}. A further challenge of ML in healthcare is its potential for poor generalizability \cite{liu_medical_2022}. To mitigate this issue, the authors recommend a medical algorithmic audit, which involves both users and developers as joint stakeholders. In our study, we addressed this challenge by working closely with surgeons.

\section{Conclusion}
\label{sec:conclusion}

In our text-to-image approach, we analyze three diffusion-based model architectures, which generate synthetic laparoscopic images from text-prompts. Their quality is assessed by comparing the human perception of produced samples and their added value in a recognition task. On this basis, we derive three key conclusions. First, our approach works best with the diffusion-based Imagen and Elucidated Imagen models, due to their superior performance in terms of fidelity and diversity compared to the Dall-e2 model. Second, our study shows that laparoscopic samples have the potential to enhance recognition performance of CV-enabled surgical applications. Third, humans with medical backgrounds often perceive our samples as actual laparoscopic images.

A challenge for future work is the need for more precise image descriptions to generate more accurate and relevant images, which could allow for the generation of synthetic laparoscopic videos. Recent diffusion-based models already generated videos in other domains that appear realistic \cite{ho_video_2022}. This is particularly interesting in order to generate dynamic and interactive surgical simulations. A promising field of research lies ahead.

\bibliographystyle{unsrtnat}
\bibliography{arxiv.bib}  %

\begin{thebibliography}{34}
\providecommand{\natexlab}[1]{#1}
\providecommand{\url}[1]{\texttt{#1}}
\expandafter\ifx\csname urlstyle\endcsname\relax
  \providecommand{\doi}[1]{doi: #1}\else
  \providecommand{\doi}{doi: \begingroup \urlstyle{rm}\Url}\fi

\bibitem[Mascagni et~al.(2022)Mascagni, Alapatt, Sestini, Altieri, Madani, Watanabe, Alseidi, Redan, Alfieri, Costamagna, Boškoski, Padoy, and Hashimoto]{mascagni_computer_2022}
Pietro Mascagni, Deepak Alapatt, Luca Sestini, Maria~S. Altieri, Amin Madani, Yusuke Watanabe, Adnan Alseidi, Jay~A. Redan, Sergio Alfieri, Guido Costamagna, Ivo Boškoski, Nicolas Padoy, and Daniel~A. Hashimoto.
\newblock Computer vision in surgery: from potential to clinical value.
\newblock \emph{npj Digital Medicine}, 5\penalty0 (1):\penalty0 1--9, 2022.

\bibitem[Mazer et~al.(2022)Mazer, Varban, Montgomery, Awad, and Schulman]{mazer_video_2022}
Laura Mazer, Oliver Varban, John~R. Montgomery, Michael~M. Awad, and Allison Schulman.
\newblock Video is better: why aren’t we using it? {A} mixed-methods study of the barriers to routine procedural video recording and case review.
\newblock \emph{Surgical Endoscopy}, 36\penalty0 (2):\penalty0 1090--1097, 2022.

\bibitem[Chambon et~al.(2023)Chambon, Bluethgen, Langlotz, and Chaudhari]{chambon_adapting_2023}
Pierre Joseph~Marcel Chambon, Christian Bluethgen, Curtis Langlotz, and Akshay Chaudhari.
\newblock Adapting {Pretrained} {Vision}-{Language} {Foundational} {Models} to {Medical} {Imaging} {Domains}.
\newblock In \emph{Surgical Innovation NeurIPS Workshops}, 2023.

\bibitem[Kondo(2021)]{kondo_lapformer_2021}
Satoshi Kondo.
\newblock {LapFormer}: surgical tool detection in laparoscopic surgical video using transformer architecture.
\newblock \emph{Comput. Methods. Biomech. Biomed. Eng. Imaging Vis.}, 9\penalty0 (3):\penalty0 302--307, 2021.

\bibitem[Portelli et~al.(2020)Portelli, Bianco, Bezzina, and Abela]{portelli_virtual_2020}
M~Portelli, SF~Bianco, T~Bezzina, and JE~Abela.
\newblock Virtual reality training compared with apprenticeship training in laparoscopic surgery: a meta-analysis.
\newblock \emph{Ann. R. Coll. Surg. Engl.}, 102\penalty0 (9):\penalty0 672--684, 2020.

\bibitem[Jin et~al.(2021)Jin, Dai, and Wang]{jin_application_2021}
Chi Jin, Liuyan Dai, and Tong Wang.
\newblock The application of virtual reality in the training of laparoscopic surgery: {A} systematic review and meta-analysis.
\newblock \emph{IJS}, 87:\penalty0 105859, 2021.

\bibitem[Pfeiffer et~al.(2019)Pfeiffer, Funke, Robu, Bodenstedt, Strenger, Engelhardt, Roß, Clarkson, Gurusamy, Davidson, Maier-Hein, Riediger, Welsch, Weitz, and Speidel]{pfeiffer_generating_2019}
Micha Pfeiffer, Isabel Funke, Maria~R. Robu, Sebastian Bodenstedt, Leon Strenger, Sandy Engelhardt, Tobias Roß, Matthew~J. Clarkson, Kurinchi Gurusamy, Brian~R. Davidson, Lena Maier-Hein, Carina Riediger, Thilo Welsch, Jürgen Weitz, and Stefanie Speidel.
\newblock Generating {Large} {Labeled} {Data} {Sets} for {Laparoscopic} {Image} {Processing} {Tasks} {Using} {Unpaired} {Image}-to-{Image} {Translation}.
\newblock In Dinggang Shen, Tianming Liu, Terry~M. Peters, Lawrence~H. Staib, Caroline Essert, Sean Zhou, Pew-Thian Yap, and Ali Khan, editors, \emph{{MICCAI}}, Cham, 2019.

\bibitem[Wu et~al.(2023)Wu, Yin, Qi, Wang, Tang, and Duan]{wu2023visual}
Chenfei Wu, Shengming Yin, Weizhen Qi, Xiaodong Wang, Zecheng Tang, and Nan Duan.
\newblock Visual chatgpt: Talking, drawing and editing with visual foundation models, 2023.

\bibitem[Sharan et~al.(2022)Sharan, Romano, Koehler, Kelm, Karck, De~Simone, and Engelhardt]{sharan_mutually_2022}
Lalith Sharan, Gabriele Romano, Sven Koehler, Halvar Kelm, Matthias Karck, Raffaele De~Simone, and Sandy Engelhardt.
\newblock Mutually improved endoscopic image synthesis and landmark detection in unpaired image-to-image translation.
\newblock \emph{JBHI}, 26\penalty0 (1):\penalty0 127--138, 2022.

\bibitem[Marzullo et~al.(2021)Marzullo, Moccia, Catellani, Calimeri, and Momi]{marzullo_towards_2021}
Aldo Marzullo, Sara Moccia, Michele Catellani, Francesco Calimeri, and Elena~De Momi.
\newblock Towards realistic laparoscopic image generation using image-domain translation.
\newblock \emph{Comput. Methods Programs Biomed.}, 200:\penalty0 105834, 2021.

\bibitem[Colleoni and Stoyanov(2021)]{colleoni_robotic_2021}
Emanuele Colleoni and Danail Stoyanov.
\newblock Robotic {Instrument} {Segmentation} {With} {Image}-to-{Image} {Translation}.
\newblock \emph{IEEE RA-L}, 6\penalty0 (2):\penalty0 935--942, 2021.

\bibitem[Pinaya et~al.(2022)Pinaya, Tudosiu, Dafflon, Da~Costa, Fernandez, Nachev, Ourselin, and Cardoso]{pinaya_brain_2022}
Walter H.~L. Pinaya, Petru-Daniel Tudosiu, Jessica Dafflon, Pedro~F. Da~Costa, Virginia Fernandez, Parashkev Nachev, Sebastien Ourselin, and M.~Jorge Cardoso.
\newblock Brain {Imaging} {Generation} with {Latent} {Diffusion} {Models}.
\newblock In Anirban Mukhopadhyay, Ilkay Oksuz, Sandy Engelhardt, Dajiang Zhu, and Yixuan Yuan, editors, \emph{DGM4MICCAI}, Cham, 2022.

\bibitem[Ali et~al.(2023)Ali, Murad, and Shah]{ali_spot_2023}
Hazrat Ali, Shafaq Murad, and Zubair Shah.
\newblock Spot the {Fake} {Lungs}: {Generating} {Synthetic} {Medical} {Images} {Using} {Neural} {Diffusion} {Models}.
\newblock In Luca Longo and Ruairi O’Reilly, editors, \emph{AICS}, 2023.

\bibitem[Sagers et~al.(2023)Sagers, Diao, Groh, Rajpurkar, Adamson, and Manrai]{sagers_improving_2023}
Luke~William Sagers, James~Allen Diao, Matthew Groh, Pranav Rajpurkar, Adewole Adamson, and Arjun~Kumar Manrai.
\newblock Improving dermatology classifiers across populations using images generated by large diffusion models.
\newblock In \emph{{NeurIPS} {Workshops}}, 2023.

\bibitem[Karras et~al.(2022)Karras, Aittala, Aila, and Laine]{karras_elucidating_2022}
Tero Karras, Miika Aittala, Timo Aila, and Samuli Laine.
\newblock Elucidating the {Design} {Space} of {Diffusion}-{Based} {Generative} {Models}.
\newblock In \emph{NeurIPS}, 2022.

\bibitem[Ramesh et~al.(2022)Ramesh, Dhariwal, Nichol, Chu, and Chen]{ramesh_hierarchical_2022}
Aditya Ramesh, Prafulla Dhariwal, Alex Nichol, Casey Chu, and Mark Chen.
\newblock Hierarchical {Text}-{Conditional} {Image} {Generation} with {CLIP} {Latents}, 2022.

\bibitem[Saharia et~al.(2022)Saharia, Chan, Saxena, Li, Whang, Denton, Ghasemipour, Gontijo-Lopes, Ayan, Salimans, Ho, Fleet, and Norouzi]{saharia_photorealistic_2022}
Chitwan Saharia, William Chan, Saurabh Saxena, Lala Li, Jay Whang, Emily Denton, Seyed Kamyar~Seyed Ghasemipour, Raphael Gontijo-Lopes, Burcu~Karagol Ayan, Tim Salimans, Jonathan Ho, David~J. Fleet, and Mohammad Norouzi.
\newblock Photorealistic {Text}-to-{Image} {Diffusion} {Models} with {Deep} {Language} {Understanding}.
\newblock In \emph{NeurIPS}, 2022.

\bibitem[Dhariwal and Nichol(2021)]{dhariwal_diffusion_2021}
Prafulla Dhariwal and Alexander Nichol.
\newblock Diffusion {Models} {Beat} {GANs} on {Image} {Synthesis}.
\newblock In \emph{NeurIPS 2021}, 2021.

\bibitem[Heusel et~al.(2017)Heusel, Ramsauer, Unterthiner, Nessler, and Hochreiter]{heusel_gans_2017}
Martin Heusel, Hubert Ramsauer, Thomas Unterthiner, Bernhard Nessler, and Sepp Hochreiter.
\newblock {GANs} {Trained} by a {Two} {Time}-{Scale} {Update} {Rule} {Converge} to a {Local} {Nash} {Equilibrium}.
\newblock In \emph{NeurIPS}, 2017.

\bibitem[Betzalel et~al.(2022)Betzalel, Penso, Navon, and Fetaya]{betzalel_study_2022}
Eyal Betzalel, Coby Penso, Aviv Navon, and Ethan Fetaya.
\newblock A {Study} on the {Evaluation} of {Generative} {Models}, 2022.

\bibitem[Parmar et~al.(2022)Parmar, Zhang, and Zhu]{parmar_aliased_2022}
Gaurav Parmar, Richard Zhang, and Jun-Yan Zhu.
\newblock On {Aliased} {Resizing} and {Surprising} {Subtleties} in {GAN} {Evaluation}.
\newblock In \emph{{CVPR}}, 2022.

\bibitem[Radford et~al.(2021)Radford, Kim, Hallacy, Ramesh, Goh, Agarwal, Sastry, Askell, Mishkin, Clark, Krueger, and Sutskever]{radford_learning_2021}
Alec Radford, Jong~Wook Kim, Chris Hallacy, A.~Ramesh, Gabriel Goh, Sandhini Agarwal, Girish Sastry, Amanda Askell, Pamela Mishkin, Jack Clark, Gretchen Krueger, and Ilya Sutskever.
\newblock Learning {Transferable} {Visual} {Models} {From} {Natural} {Language} {Supervision}.
\newblock In \emph{ICML}, 2021.

\bibitem[Raffel et~al.(2020)Raffel, Shazeer, Roberts, Lee, Narang, Matena, Zhou, Li, and Liu]{raffel_exploring_2020}
Colin Raffel, Noam Shazeer, Adam Roberts, Katherine Lee, Sharan Narang, Michael Matena, Yanqi Zhou, Wei Li, and Peter~J. Liu.
\newblock Exploring the limits of transfer learning with a unified text-to-text transformer.
\newblock \emph{JMLR}, 21\penalty0 (1):\penalty0 140:5485--140:5551, 2020.

\bibitem[Choi et~al.(2022)Choi, Lee, Shin, Kim, Kim, and Yoon]{choi_perception_2022}
Jooyoung Choi, Jungbeom Lee, Chaehun Shin, Sungwon Kim, Hyunwoo Kim, and Sungroh Yoon.
\newblock Perception {Prioritized} {Training} of {Diffusion} {Models}.
\newblock In \emph{{CVPRW}}, 2022.

\bibitem[Nwoye et~al.(2022)Nwoye, Yu, Gonzalez, Seeliger, Mascagni, Mutter, Marescaux, and Padoy]{nwoye_rendezvous_2022}
Chinedu~Innocent Nwoye, Tong Yu, Cristians Gonzalez, Barbara Seeliger, Pietro Mascagni, Didier Mutter, Jacques Marescaux, and Nicolas Padoy.
\newblock Rendezvous: {Attention} mechanisms for the recognition of surgical action triplets in endoscopic videos.
\newblock \emph{Medical Image Analysis}, 78:\penalty0 102433, 2022.

\bibitem[Twinanda et~al.(2017)Twinanda, Shehata, Mutter, Marescaux, de~Mathelin, and Padoy]{twinanda_endonet_2017}
Andru~P Twinanda, Sherif Shehata, Didier Mutter, Jacques Marescaux, Michel de~Mathelin, and Nicolas Padoy.
\newblock {EndoNet}: {A} {Deep} {Architecture} for {Recognition} {Tasks} on {Laparoscopic} {Videos}.
\newblock \emph{TMI}, 36\penalty0 (1):\penalty0 86--97, 2017.

\bibitem[Hong et~al.(2020)Hong, Kao, Kuo, Wang, Chang, and Shih]{hong_cholecseg8k_2020}
W.-Y. Hong, C.-L. Kao, Y.-H. Kuo, J.-R. Wang, W.-L. Chang, and C.-S. Shih.
\newblock {CholecSeg8k}: {A} {Semantic} {Segmentation} {Dataset} for {Laparoscopic} {Cholecystectomy} {Based} on {Cholec80}, 2020.

\bibitem[Naeem et~al.(2020)Naeem, Oh, Uh, Choi, and Yoo]{naeem_reliable_2020}
Muhammad~Ferjad Naeem, Seong~Joon Oh, Youngjung Uh, Yunjey Choi, and Jaejun Yoo.
\newblock Reliable {Fidelity} and {Diversity} {Metrics} for {Generative} {Models}.
\newblock In \emph{ICML}, 2020.

\bibitem[Bińkowski et~al.(2023)Bińkowski, Sutherland, Arbel, and Gretton]{binkowski_demystifying_2023}
Mikołaj Bińkowski, Danica~J. Sutherland, Michael Arbel, and Arthur Gretton.
\newblock Demystifying {MMD} {GANs}.
\newblock In \emph{ICLR}, 2023.

\bibitem[Nwoye and Padoy(2023)]{nwoye_data_2023}
Chinedu~Innocent Nwoye and Nicolas Padoy.
\newblock Data {Splits} and {Metrics} for {Method} {Benchmarking} on {Surgical} {Action} {Triplet} {Datasets}, 2023.

\bibitem[Carlini et~al.(2023)Carlini, Hayes, Nasr, Jagielski, Sehwag, Tram{\`e}r, Balle, Ippolito, and Wallace]{carlini_extracting_2023}
Nicolas Carlini, Jamie Hayes, Milad Nasr, Matthew Jagielski, Vikash Sehwag, Florian Tram{\`e}r, Borja Balle, Daphne Ippolito, and Eric Wallace.
\newblock Extracting training data from diffusion models.
\newblock In \emph{32nd USENIX Security Symposium (USENIX Security 23)}, 2023.

\bibitem[Markus et~al.(2021)Markus, Kors, and Rijnbeek]{markus_role_2021}
Aniek~F. Markus, Jan~A. Kors, and Peter~R. Rijnbeek.
\newblock The role of explainability in creating trustworthy artificial intelligence for health care: {A} comprehensive survey of the terminology, design choices, and evaluation strategies.
\newblock \emph{JBI}, 113:\penalty0 103655, 2021.

\bibitem[Liu et~al.(2022)Liu, Glocker, McCradden, Ghassemi, Denniston, and Oakden-Rayner]{liu_medical_2022}
Xiaoxuan Liu, Ben Glocker, Melissa~M McCradden, Marzyeh Ghassemi, Alastair~K Denniston, and Lauren Oakden-Rayner.
\newblock The medical algorithmic audit.
\newblock \emph{The Lancet Digital Health}, 4\penalty0 (5):\penalty0 e384--e397, 2022.

\bibitem[Ho et~al.(2022)Ho, Salimans, Gritsenko, Chan, Norouzi, and Fleet]{ho_video_2022}
Jonathan Ho, Tim Salimans, Alexey Gritsenko, William Chan, Mohammad Norouzi, and David~J. Fleet.
\newblock Video {Diffusion} {Models}, 2022.

\end{thebibliography}

\end{document}